\documentclass[a4paper,11pt]{article}

\usepackage{stmaryrd}
\usepackage{bm}
\usepackage{cite}
\usepackage{color}
\usepackage{dsfont}
\usepackage{nicefrac}
\usepackage{multirow}
\usepackage{graphics}
\usepackage{graphicx}
\usepackage{epsfig}
\usepackage{subfigure}
\usepackage{pgf}
\usepackage{amssymb}
\usepackage{amsmath,amssymb,amsfonts}
\usepackage{wasysym}

\begin{document}

\title{Inverse design of resonant nanostructures for extraordinary optical transmission of periodic metallic slits}

\author{Yongbo Deng$^1$\footnote{dengyb@ciomp.ac.cn}, Chao Song$^2$, Yihui Wu$^1$, Zhenyu Liu$^1$, Jan G. Korvink$^3$\\
$^1$ State Key Laboratory of Applied Optics, \\
Changchun Institute of Optics, Fine Mechanics and Physics (CIOMP), \\
Chinese Academy of Sciences, Changchun, Jilin 130033, China \\
$^2$ Department of Electrical Engineering, Iowa State University \\
$^3$ Institute of Microstructure Technology (IMT), \\
Karlsruhe Institute of Technology (KIT), \\
Hermann-von-Helmholtzplatz 1, Eggenstein-Leopoldshafen 76344, Germany}
\maketitle

\abstract{This paper presents the inverse design of resonant nanostructures for the extraordinary optical transmission of periodic metallic slits, where the topology optimization approach is utilized to implement the inverse design procedure and find the geometrical configurations of the nanostructures. By the inverse design method, the subwavelength-size resonant nanostructures, localized at the inlet and outlet sides of the periodic metallic slits, are derived with transmission peaks at the prescribed incident wavelengths. And the transmissivity is enhanced by effective excitation and guidance of surface plasmon polariton at the inlet side of the slits, coherently resonance of surface plasmon polariton inside the slits and radiation of the photonic energy at the outlet side of the slits. The transmission peaks of the periodic metallic slits, with inversely designed resonant nanostructures, are raised along with the red shift of the incident wavelength. The presentation of the transmission peak of periodic metallic slits can be controlled and localized at a desired frequency, by specifying the incident wave with the wavelength corresponding to the desired frequency for the inverse design procedure. By maximizing the minimum transmissivity of the periodic metallic slits with incident wavelengths in a prescribed wavelength range, the extraordinary optical transmission bandwidth can be enlarged, and the sensitivity of transmissivity to wavelength can be decreased equivalently.

}

\section{Introduction}

Extraordinary optical transmission (EOT) is the phenomenon of greatly enhanced transmission of light through a subwavelength aperture in an otherwise opaque metallic film which has been patterned with a regularly repeating periodic structure. It was first described by Ebbesen et al in 1998 \cite{Ebbesen1998}. In EOT, the regularly repeating structures enable much higher transmissivity to occur, up to several orders of magnitude greater than that predicted by classical aperture theory. And the mechanism of EOT is attributed to the scattering of surface plasmon polaritons (SPPs) \cite{Liu2008,Abajo2007}. EOT offers one key advantage over a surface plasmonic resonance (SPR) device, an inherently nanometer-micrometer scale device, and it is particularly amenable to miniaturization. Tremendous potential applications of EOT include several newly emerging areas, e.g. subwavelength optics, opto-electronic devices, wavelength-tunable filters, optical modulators \cite{Barnes2003,Engheta2007,Genet2007,Abajo2007}, left-handed metamaterial and chemical sensing \cite{Gordon2008}. To achieve the required transmission performance, metallic layouts with subwavelength apertures, e.g. subwavelength hole array \cite{Ebbesen1998}, periodic slit array, tapered slits \cite{Sodergaard2010}, diatomic chain of slit-hole \cite{Liu2009}, groove array flanking slit \cite{Garcia-Vidal2003} and bull's eye structures\cite{Wang2009}, have been proposed for EOT; and parametric optimization of metallic layouts with subwavelength apertures have been implemented to enhance the transmissivity \cite{Cui2008,Popov2005}. Most of these researches are focused on enhancing EOT with nanostructures derived based on physical intuition. The more challenging problem problem can be inverse design or determination of the geometrical configurations of the nanostructures for enhancing EOT. Therefore, this paper is devoted to inversely designing resonant nanostructures to enhance the extraordinary optical transmission of periodic metallic slits, where the topology optimization approach is utilized to implement the inverse design method and find the geometrical configurations of the nanostructures.

Recently, it is shown that topology optimization can be used to inversely determine geometrical configuration of structures and achieve the inverse design of devices in elastics, hydrodynamics, electromagnetics or photonics et al \cite{Bendsoe1990,Borrvall2003,Jensen2005,Bendsoe2003}. In electromagnetics or photonics, topology optimization approach has been applied in the inverse design of beamsplitters \cite{Piggott2015,Shen2015}, photonic crystals \cite{Sigmund2008,Frandsen2014,Men2014}, cloaks \cite{Andkjaer2011,Andkjaer2012,Fujii2013}, metamaterials \cite{Otomori2012,Diaz2010,Zhou2011}, excitation of SPPs and LSPRs \cite{Andkjaer20101,Deng2015}, and electromagnetic and optical antennas \cite{Zhou20102,Feichtner2012,Hassan2014,Erentok2011}. In topology optimization, the structure is inversely determined using the material penalization approach, where the design variable is used to represent the material distribution and geometrical configuration. The design variable is evolved to a indicator function using the robust gradient based optimization algorithm, e.g. the method of moving asymptotes (MMA) \cite{Svanberg1987}. Therefore, the topology optimization approach is chosen to implement the inverse design of the nanostructures of the periodic metallic slits for EOT.

\section{Methodology}

\begin{figure}[!htbp]
  \centering
  \includegraphics[width=0.7\columnwidth]{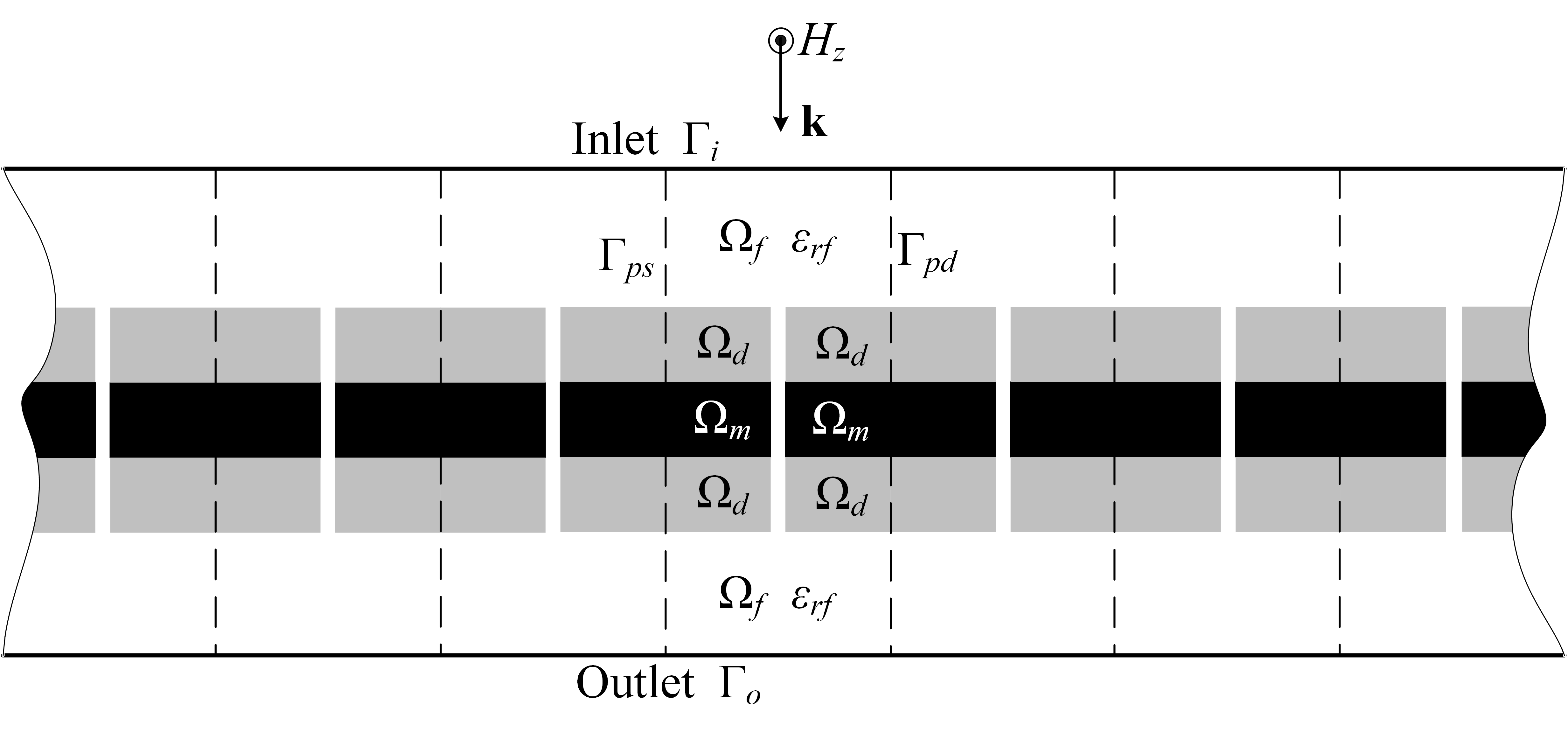}\\
  \caption{Computational domain for the inverse design of bilateral nanostructures for the periodic metallic slits, where $H_z$ is the propagating TM wave; $\mathbf{k}$ is the wave vector; $\Omega_f$, $\Omega_d$ and $\Omega_m$ are the free space, design and metallic domains, respectively; $\Gamma_i$ and $\Gamma_o$ are inlet and outlet boundaries of the photonic energy, respectively; $\epsilon_{rf}$, $\epsilon_{rd}$ and $\epsilon_{rm}$ respectively are the relative permittivity in $\Omega_f$, $\Omega_d$ and $\Omega_m$; $\Gamma_{ps}$ and $\Gamma_{pd}$ are the source and destination boundary of the periodic boundary pair, respectively; $\Omega=\Omega_f\cup\Omega_d\cup\Omega_m$ is the computational domain.}\label{CompDom}
\end{figure}

Cross-section of periodic metallic slits with infinite thickness is illuminated in Fig. \ref{CompDom} with a uniform monochromatic transverse magnetic (TM) wave propagation. The computational domain is set to be one period of the metallic slits. The topology optimization approach is utilized to inversely design the nanostructures localized in the bilateral regions of the preset metallic slits. Then the design domain, where the design variable is defined, is set to be these two bilateral regions. To truncate the infinitive free space, the first order absorbing boundary condition is imposed on the inlet ($\Gamma_i$) and outlet ($\Gamma_o$) boundaries of the computational domain, and the periodic boundary condition is imposed on the left ($\Gamma_{ps}$) and right ($\Gamma_{pd}$) boundaries of the slit to reduce the computational cost. Based on the above computational setup, the inverse design problem is to find the geometrical configurations of the bilateral nanostructures for the preset slit to maximize the transmission of the electromagnetic energy. The propagating wave in the metallic slits is time-harmonic TM wave governed by the Maxwell's equations in two dimensions, which can be reformulated into the scalar Helmholtz equation expressed as
\begin{equation}\label{WaveEquHz}
\begin{split}
    & \nabla\cdot \left[\epsilon_r^{-1}\nabla \left(H_{zs} + H_{zi}\right)\right] + k_0^2 \mu_r \left(H_{zs} + H_{zi}\right) = 0,~\mathrm{in}~\Omega\\
    & \epsilon_r^{-1} \nabla H_{zs} \cdot \mathbf{n} + j k_0 \sqrt{\epsilon_r^{-1}\mu_r} H_{zs} = 0,~\mathrm{on}~\Gamma_i\cup\Gamma_o \\
    & H_{zs}\left(\mathbf{x}+\mathbf{a}\right) = H_{zs}\left(\mathbf{x}\right)e^{-j\mathbf{k}\cdot\mathbf{a}},~\mathbf{n}\left(\mathbf{x}+\mathbf{a}\right)\cdot\nabla H_{zs}\left(\mathbf{x}+\mathbf{a}\right) = - e^{-j\mathbf{k}\cdot\mathbf{a}} \mathbf{n}\left(\mathbf{x}\right)\cdot\nabla H_{zs}\left(\mathbf{x}\right),\\
    & \mathrm{for}~\forall \mathbf{x}\in\Gamma_{ps},~\mathbf{x}+\mathbf{a}\in\Gamma_{pd}
\end{split}
\end{equation}
where the scattered-field formulation, with $H_z = H_{zs} + H_{zi}$, is used to reduce the dispersion
error; $H_{zs}$ and $H_{zi}$ are the scattered and incident fields, respectively; $\epsilon_r$ and $\mu_r$ are the relative permittivity and permeability, respectively; $k_0 = \omega\sqrt{\epsilon_0 \mu_0}$ is the free space wave number with $\omega$, $\epsilon_0$ and $\mu_0$ respectively representing the angular frequency, free space permittivity and permeability; $\Omega$ is the computational domain; $\mathbf{k}$ is the wave vector; the time dependence of the fields is given by the factor $e^{j\omega t}$, with $t$ representing the time; $\mathbf{n}$ is the unit outward normal vector at $\partial\Omega$; $j=\sqrt{-1}$ is the imaginary unit; $\Gamma_i$ and $\Gamma_o$ are respectively the inlet and outlet boundaries of the photonic energy; and $\Gamma_{ps}$ and $\Gamma_{pd}$ are respectively the source and destination boundary of the periodic boundary pair, with lattice vector $\mathbf{a}$. The incident field $H_{zi}$ is set to be the parallel-plane wave with unit amplitude.

The topology optimization approach, chosen to carry out inverse design of bilateral nanostructures for enhancing EOT of periodic metallic slits, is implemented based on the material interpolation between the metal and free space in the design domain. In EOT, the used noble metal is usually nonmagnetic, e.g. silver (Ag) and gold (Au). Therefore, the permeability is set to be 1. Then, only the spatial distribution of relative permittivity needs to be determined in the inverse design procedure. In the visible light region, the relative permittivity of noble metal can be described by the Drude model
\begin{equation}\label{Drude}
    \epsilon_{rm} = \epsilon_{r\infty} - {{\omega_p^2}\over{\omega\left(\omega-j\gamma_c\right)}}
\end{equation}
where $\epsilon_{r\infty}$ is the high-frequency bulk permittivity; $\omega$ is the angular frequency of the incident wave; $\omega_p$ is the bulk plasmon frequency; $\gamma_c$ is the collision frequency.
The material interpolation is performed using the material density representing the geometrical configuration \cite{Sigmund2006}, and the material density is introduced with the values $0$ and $1$ respectively representing free space and metal. Because the surface plasmons are presented at the surface of noble metal, the electromagnetic field decays exponentially; hence, the material density should decays rapidly away from 1 to mimic the metal surface. Thus, the material interpolation is set to be the hybrid of logarithmic and power law approaches
\begin{equation}\label{Interpolation}
    \epsilon_{rd}\left(\omega\right) = 10^{\log{\epsilon_{rm}\left(\omega\right)}-{{1-\bar{\rho}^3}\over{1+\bar{\rho}^3}}
    \left[\log{\epsilon_{rm}\left(\omega\right)}
    -\log{\epsilon_{rf}\left(\omega\right)}\right]}
\end{equation}
where $\epsilon_{rd}$ is the relative permittivity in the design domain; $\bar{\rho}$ is the material density. The material density is derived by filtering and projecting the design variable valued in $\left[0,1\right]$. During the evolution of the design variable, filtering is implemented by the density filter to enforce a minimum length scale of the derived geometrical configuration \cite{Lazarov2011}
\begin{equation}\label{HelmFilterGa}
\begin{split}
    -r^2 \nabla\cdot\nabla\tilde{\rho}+\tilde{\rho} = & ~ \rho ,~\mathrm{in}~\Omega\\
    \nabla \tilde{\rho} \cdot \mathbf{n} = & ~ 0,~\mathrm{on}~\partial \Omega
\end{split}
\end{equation}
where $r$ is the filter radius chosen based on numerical experiments \cite{Wang2011}; $\tilde{\rho}$ is the filtered design variable. And projecting is implemented by the threshold projection to remove the gray area in the derived geometrical configuration \cite{Guest2004}
\begin{equation}\label{Project}
\bar{\rho} = {{\mathrm{tanh}\left(\beta\xi\right)+\mathrm{tanh}\left(\beta\left(\tilde{\rho}-\xi\right)\right)}
     \over{\mathrm{tanh}\left(\beta\xi\right)+\mathrm{tanh}\left(\beta\left(1-\xi\right)\right)}}\\
\end{equation}
where $\xi\in\left[0,1\right]$ and $\beta$ are the threshold and projection parameters for the threshold projection, respectively. On the choice of the values of $\xi$ and $\beta$, one can refer to \cite{Xu2010,Kawamoto2010}.

EOT is featured by its high transmission of the photonic energy through periodic subwavelength metallic apertures. The input and output transmission power for one period of the metallic slits can be measured by the integration of the average Poynting vector on the the inlet and outlet sides of the computational domain
\begin{equation}\label{PowerIn}
\begin{split}
    P_i = \int_{\Gamma_i} - {1\over2} \mathrm{Re}\left(\mathbf{E}_i\times\mathbf{H}_i^*\right)\cdot\mathbf{n} \, \mathrm{d}\Gamma = \int_{\Gamma_i} \mathrm{Re}\left({1\over2j\omega\epsilon_r\epsilon_0} H_{zi}^* \nabla H_{zi}\right) \cdot\mathbf{n} \, \mathrm{d}\Gamma
\end{split}
\end{equation}
\begin{equation}\label{PowerOut}
\begin{split}
    P_o = \int_{\Gamma_o} {1\over2} \mathrm{Re}\left(\mathbf{E}\times\mathbf{H}^*\right)\cdot\mathbf{n} \, \mathrm{d}\Gamma = \int_{\Gamma_o} \mathrm{Re}\left({-1\over2j\omega\epsilon_r\epsilon_0} \left(H_{zi} + H_{zs} \right)^* \nabla \left(H_{zi} + H_{zs} \right)\right) \cdot\mathbf{n} \, \mathrm{d}\Gamma
\end{split}
\end{equation}
where $P_i$ and $P_o$ are the input and output transmission power, respectively; $\mathbf{E}_i$ is the electric fields corresponding to the incident magnetic wave $\mathbf{H}_i=\left(0,0,H_{zi}\right)$; $\mathbf{E}$ is the total electric fields; $\mathbf{H}$ is the total magnetic fields; $\mathrm{Re}$ operator is used to extract the real part of an expression; $*$ represents the conjugate of the complex variable. The objective of the inverse design procedure can be chosen to maximize the transmissivity defined as the normalized transmission power
\begin{equation}\label{Obj1}
    T_r = {P_o}/{P_i}
\end{equation}
Then the inverse design problem for the bilateral nanostructures of periodic metallic slits is to maximize the transmissivity defined in Equ. \ref{Obj1}, which is constrained by Equ. \ref{WaveEquHz} and Equ. \ref{HelmFilterGa} with design variable valued in $\left[0,1\right]$. The gradient-based iterative procedure, method of moving asymptotes \cite{Svanberg1987}, is applied to update the design variable and maximize the transmissivity. And the gradient information is obtained using the continuous adjoint method \cite{Hinze2009}
\begin{equation}\label{AdjointDerv}
\delta T_r = \int_\Omega - \mathrm{Re}\left(\tilde{\rho}_a^*\right) \delta\rho \,\mathrm{d} \Omega
\end{equation}
where $\tilde{\rho}_a$ is the adjoint of the filtered design variable $\tilde{\rho}$. $\tilde{\rho}_a$ is derived by solving the adjoint equations of the wave equation in Equ. \ref{WaveEquHz} and density filter in Equ. \ref{HelmFilterGa}
\begin{equation}\label{WeakAdjEquHz}
\begin{split}
& \int_{\Omega} - \epsilon_r^{-1} \nabla \tilde{H}_{zs}^* \cdot \nabla \phi + k_0^2 \mu_r \tilde{H}_{zs}^* \phi \,\mathrm{d}\Omega + \int_{\Gamma_o} \bigg( - j k_0 \sqrt{\epsilon_r^{-1} \mu_r} \tilde{H}_{zs}^* + {1\over{P_i}} {\partial B \over \partial H_{zs}}\bigg) \phi \,\mathrm{d}\Gamma \\
& + \int_{\Gamma_o} {1\over{P_i}} {\partial B \over \partial \nabla H_{zs}} \cdot \nabla \phi \,\mathrm{d}\Gamma = 0, ~\forall \phi \in \mathcal{H}^1\left(\Omega\right)
\end{split}
\end{equation}
\begin{equation}\label{WeakAdjEquGa}
\begin{split}
& \int_{\Omega_d} r^2 \nabla \tilde{\rho}_a^* \cdot \nabla \psi + \left[ \tilde{\rho}_a^* - {{\partial \epsilon_r^{-1}}\over{\partial \bar{\rho}}} {{\partial \bar{\rho}}\over{\partial \tilde{\rho}}} \nabla \left(H_{zs}+H_{zi}\right) \cdot \nabla \tilde{H}_{zs}^* \right] \psi \,\mathrm{d}\Omega = 0, ~\forall \psi \in \mathcal{H}^1\left(\Omega_d\right)
\end{split}
\end{equation}
where $\tilde{H}_{zs}$ and $B$ represent the adjoint of $H_{zs}$ and the integral function of Equ. \ref{PowerOut}, respectively; $\mathcal{H}^1\left(\Omega\right)$ and $\mathcal{H}^1\left(\Omega_d\right)$ are the first order Hilbert functional spaces defined on $\Omega$ and $\Omega_d$.

\section{Results and discussion}

\begin{figure}[htbp]
  \centering
  \includegraphics[height=0.34\columnwidth]{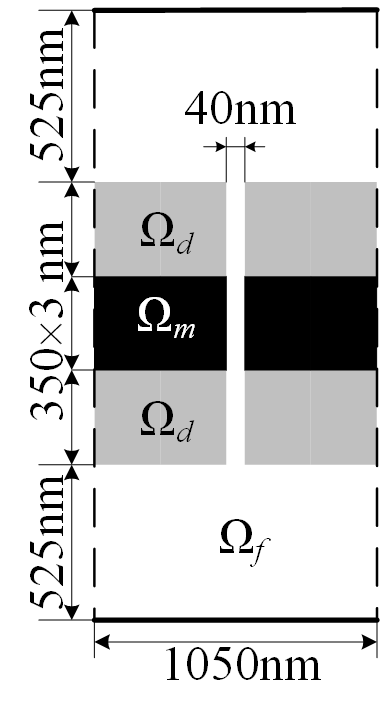}
\caption{Size presetting of the periodic metallic slits.}\label{Size}
\end{figure}

\begin{figure}[htbp]
  \centering
  \subfigure[$\lambda=350\mathrm{nm}, T_r=0.080$]
  {\includegraphics[height=0.145\columnwidth]{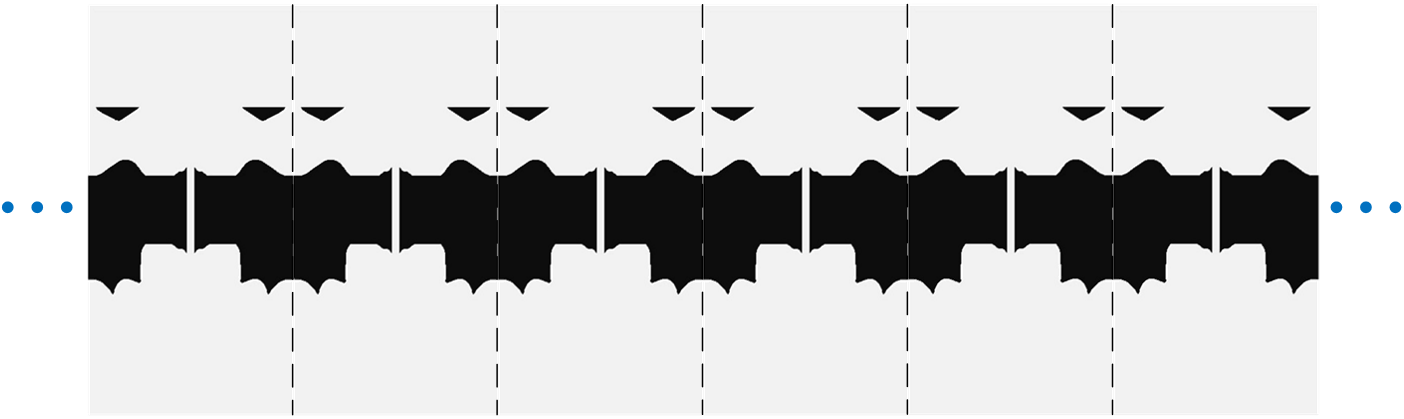}}
  \subfigure[$\lambda=440\mathrm{nm}, T_r=0.425$]
  {\includegraphics[height=0.145\columnwidth]{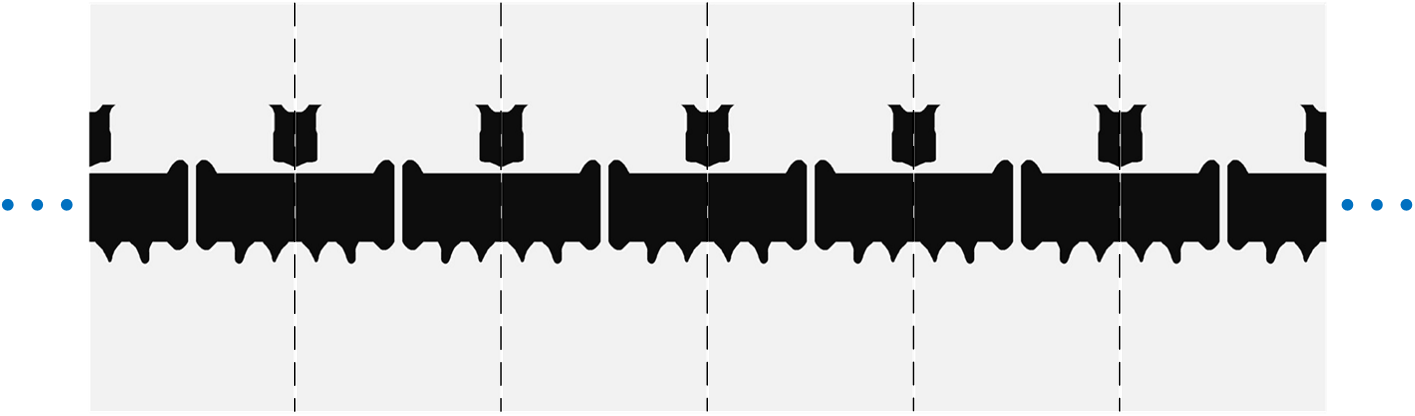}}\\
  \subfigure[$\lambda=526\mathrm{nm}, T_r=0.655$]
  {\includegraphics[height=0.145\columnwidth]{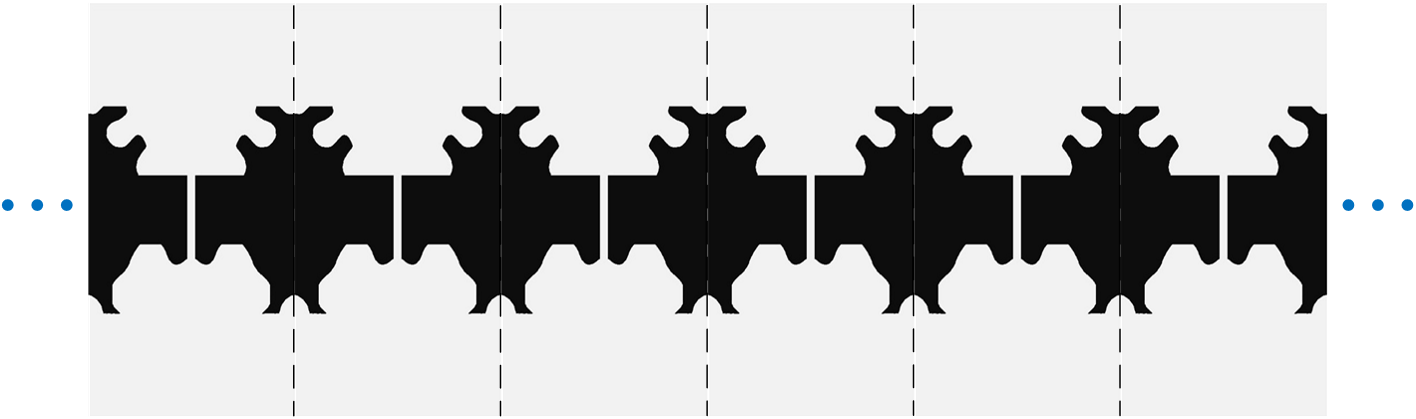}}
  \subfigure[$\lambda=616\mathrm{nm}, T_r=0.768$]
  {\includegraphics[height=0.145\columnwidth]{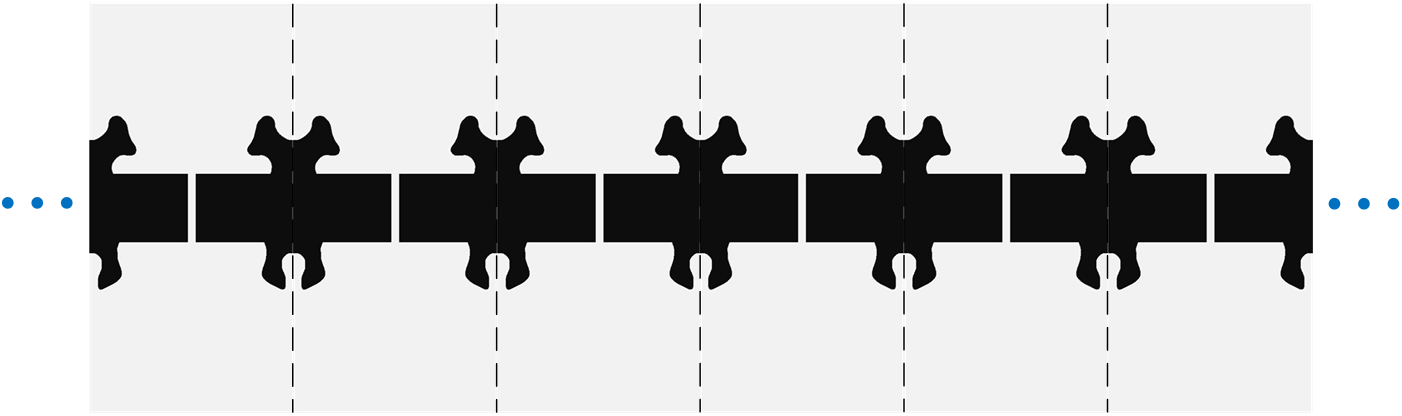}}\\
  \subfigure[$\lambda=710\mathrm{nm}, T_r=0.846$]
  {\includegraphics[height=0.145\columnwidth]{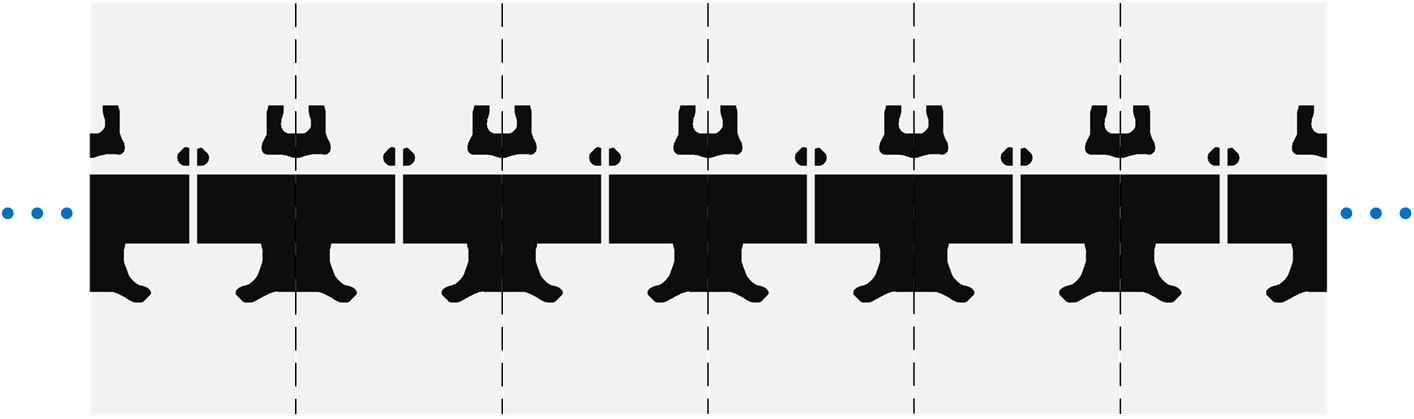}}
  \subfigure[$\lambda=770\mathrm{nm}, T_r=0.862$]
  {\includegraphics[height=0.145\columnwidth]{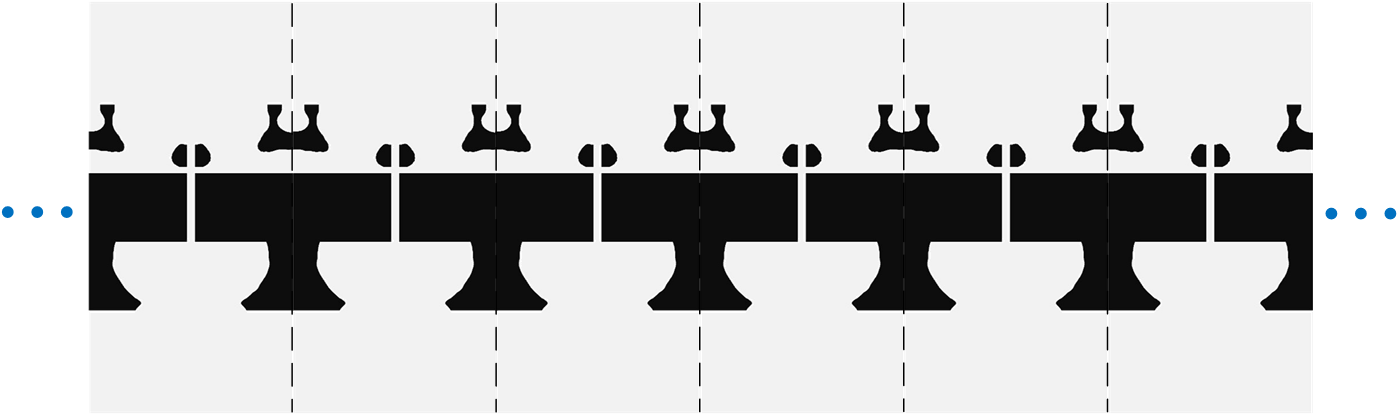}}\\
  \caption{Inversely designed bilateral nanostructures for the periodic metallic slits with extraordinary optical transmission corresponding to different incident wavelengths in the visible light region.}\label{SlitTop}
\end{figure}

\begin{figure}[htbp]
  \centering
  \subfigure[$\lambda=350\mathrm{nm}$]
  {\includegraphics[height=0.15\columnwidth]{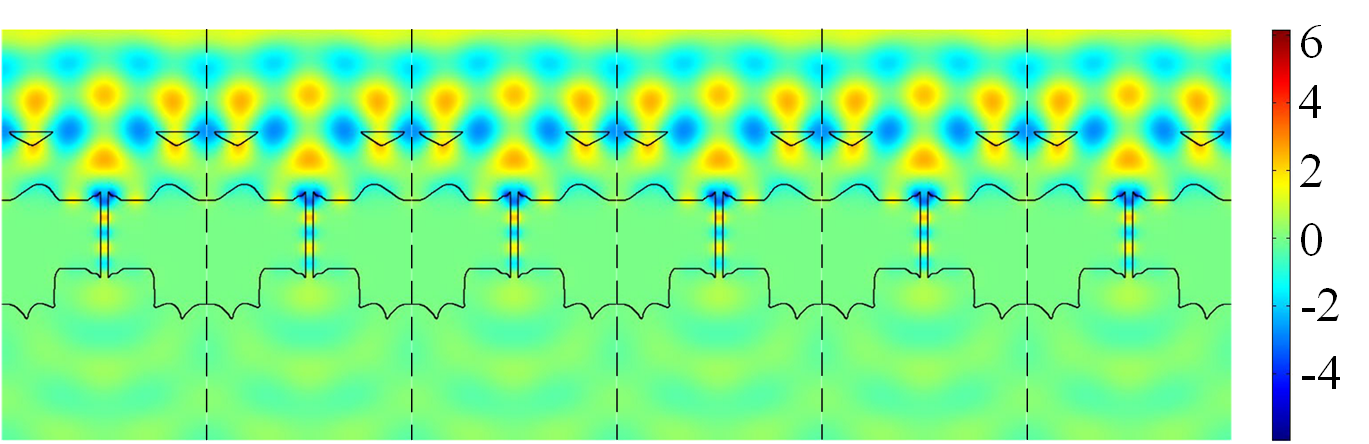}}\hspace{2em}
  \subfigure[$\lambda=440\mathrm{nm}$]
  {\includegraphics[height=0.15\columnwidth]{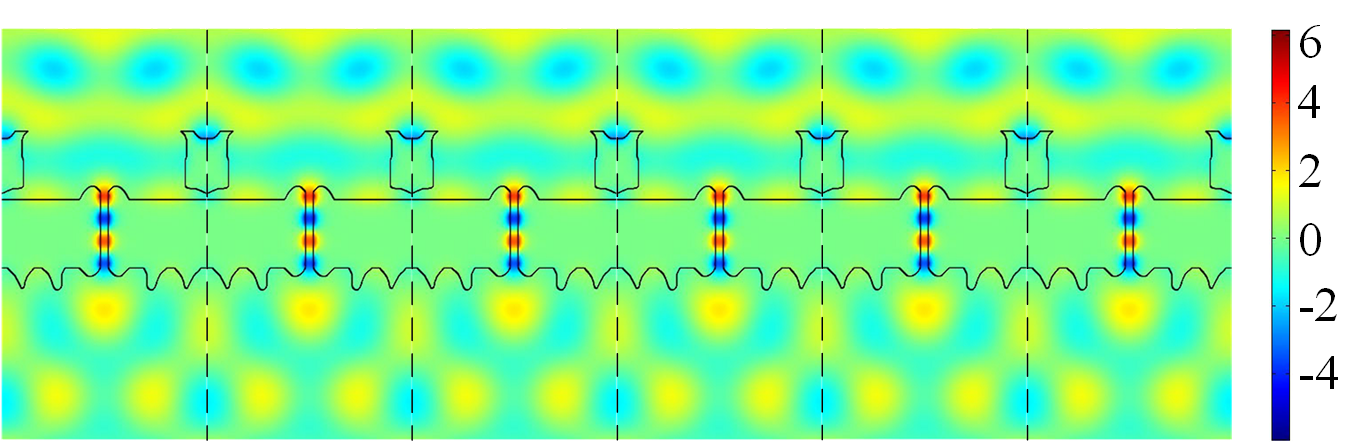}}\\
  \subfigure[$\lambda=526\mathrm{nm}$]
  {\includegraphics[height=0.15\columnwidth]{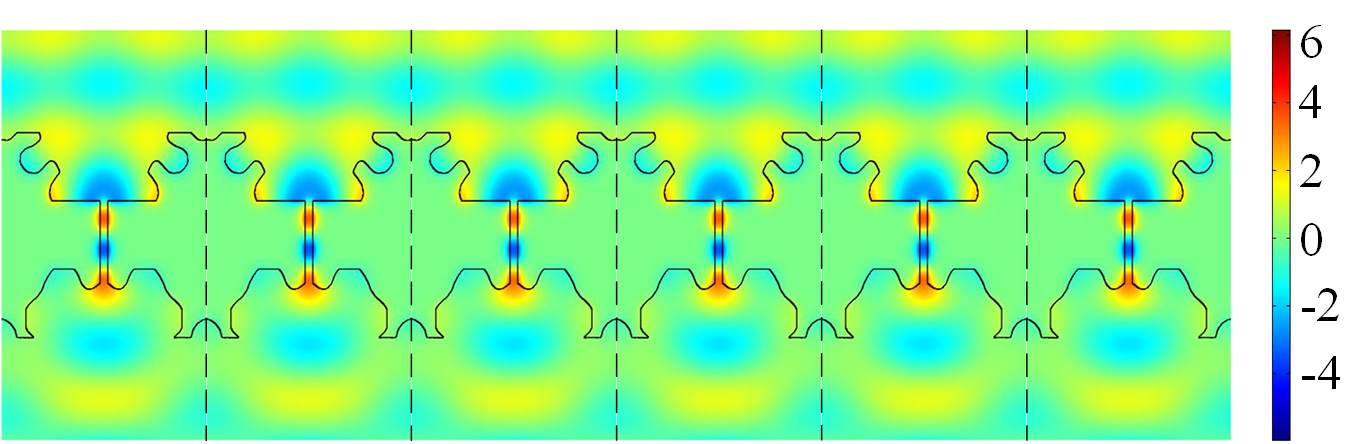}}\hspace{2em}
  \subfigure[$\lambda=616\mathrm{nm}$]
  {\includegraphics[height=0.15\columnwidth]{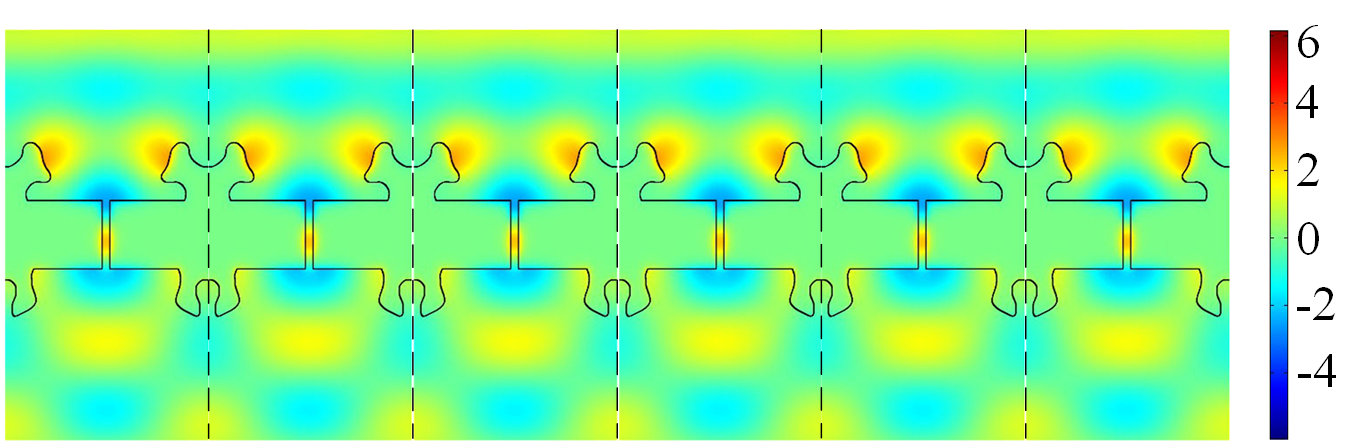}}\\
  \subfigure[$\lambda=710\mathrm{nm}$]
  {\includegraphics[height=0.15\columnwidth]{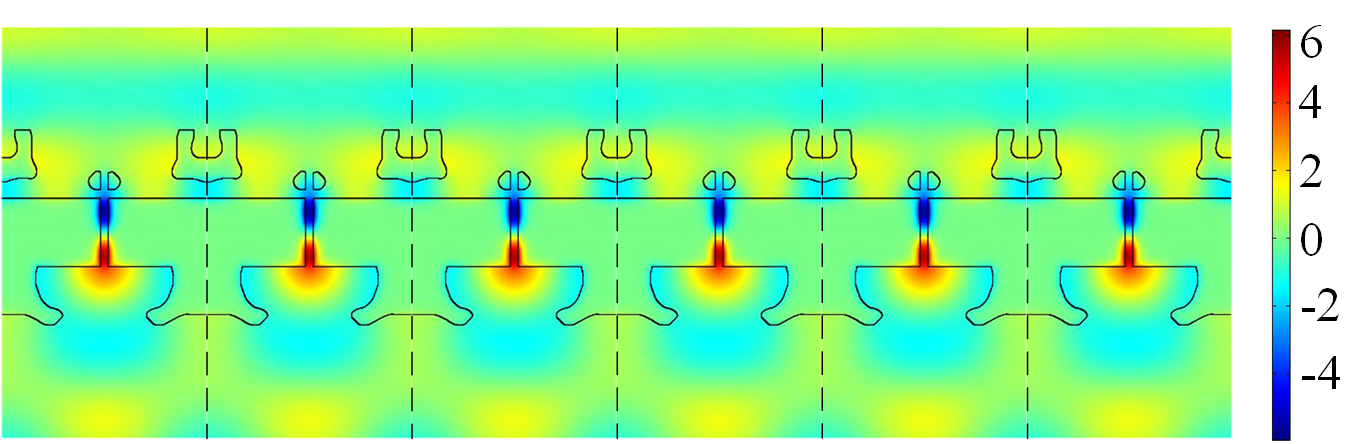}}\hspace{2em}
  \subfigure[$\lambda=770\mathrm{nm}$]
  {\includegraphics[height=0.15\columnwidth]{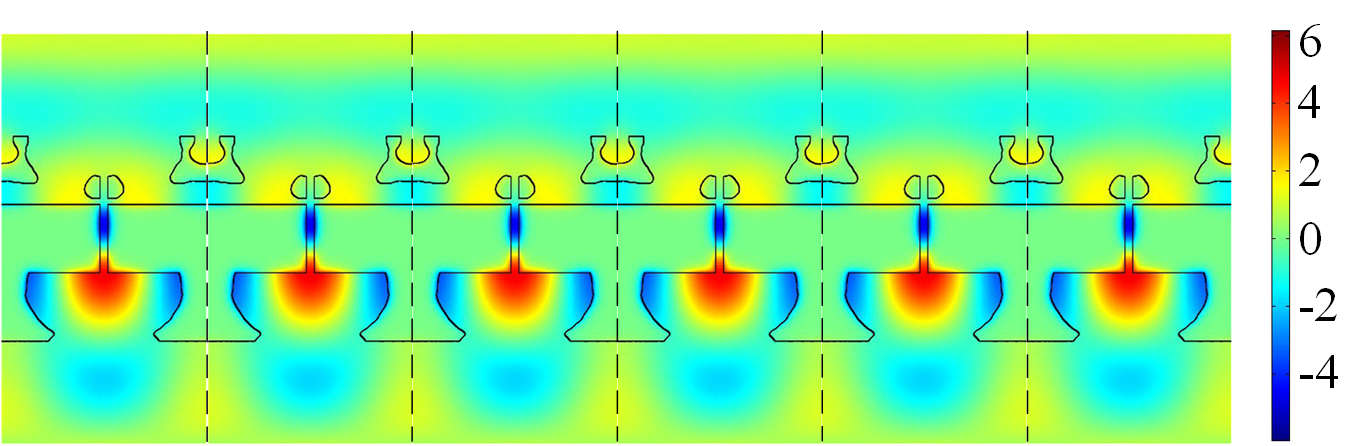}}\\
  \caption{Magnetic field distribution, in the periodic metallic slits with inversely designed bilateral nanostructures, respectively corresponding to the geometrical configurations shown in Fig. \ref{SlitTop}a$\sim$f.}\label{SlitField}
\end{figure}

\begin{figure}[htbp]
  \centering
  \includegraphics[width=0.45\columnwidth]{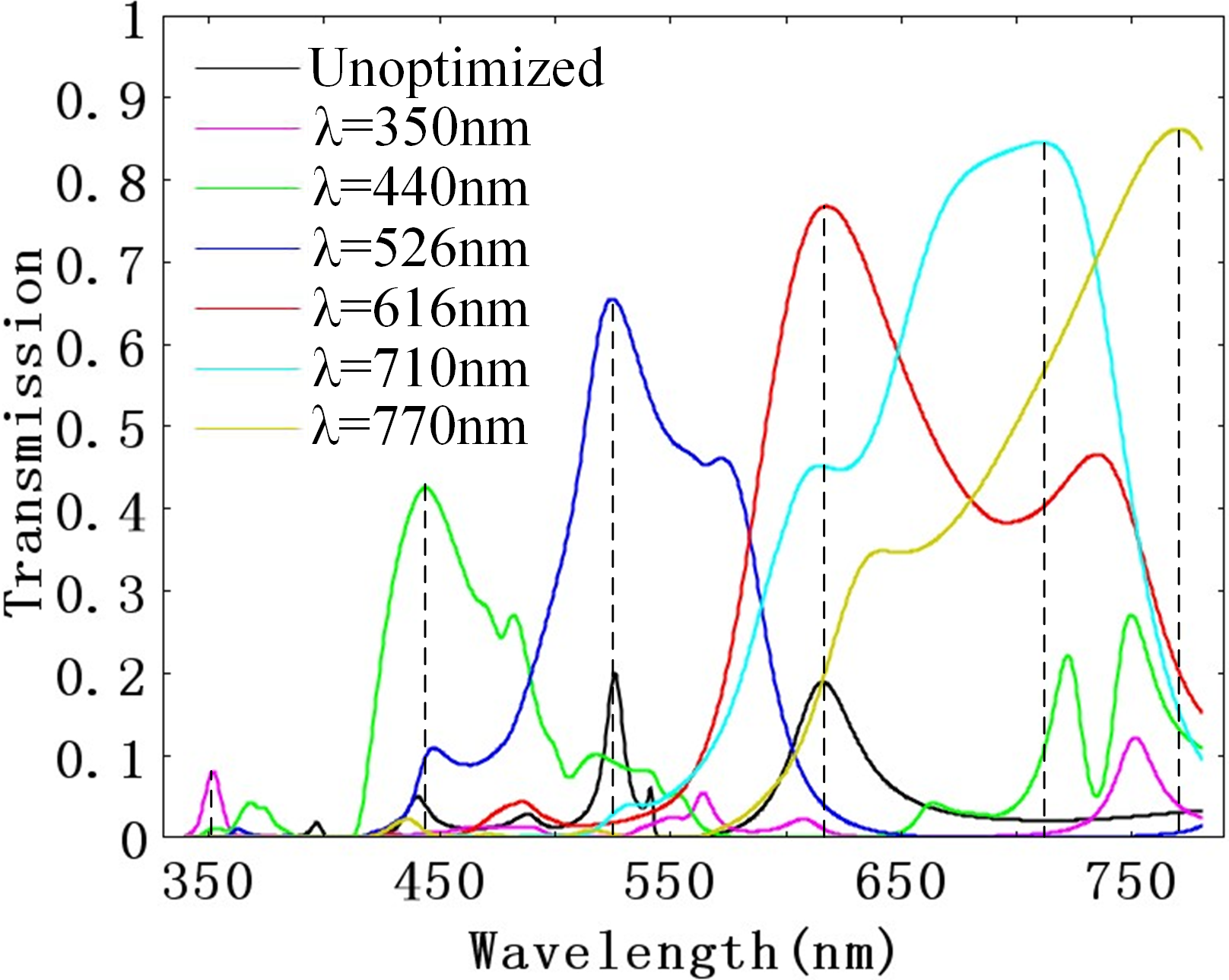}\\
  \caption{Transmission spectra of the periodic metallic slits with inversely designed bilateral nanostructures shown in Fig. \ref{SlitTop} and the periodic metallic slits without bilateral nanostructures. There are two transmission peaks at the wavelength $526$nm and $616$nm for the periodic metallic slits without bilateral nanostructures. By the inverse design method, these transmission peaks are enhanced $3.27$ and $4.06$ times, respectively. The transmission peaks of the periodic metallic slits with inversely designed bilateral nanostructures are presented at the specified incident wavelengths used in the inverse design procedure.}\label{TransSpectra}
\end{figure}

Using the outlined topology optimization based inverse design procedure for enhancing EOT of periodic metallic slits, the bilateral nanostructures are investigated as follows. The noble metal is chosen to be Ag, with high-frequency bulk permittivity $\epsilon_{r\infty}=6$, bulk plasmon frequency $\omega_p=1.5\times10^{16}$ rad/s, and collision frequency $\gamma_c=7.73\times10^{13}$ rad/s derived by fitting the experimental data in the literatures \cite{Johnson1972}. The sizes of the computational domain shown in Fig. \ref{CompDom} are set to be the typical values: $1050$nm for the periodic length of the metallic slits, $40$nm for the slit width, $350$nm for the thickness of the fixed Ag layer $\Omega_m$, and $350$nm for the thickness of the design domain $\Omega_d$ (Fig. \ref{Size}). The incident wavelength is scanned in the visible light region ($350\sim770$ nm). For different incident wavelengths, the nanostructures sticking to the metallic slits are derived as shown in Fig. \ref{SlitTop}a$\sim$f, with corresponding magnetic field distribution shown in Fig. \ref{SlitField}a$\sim$f. These results demonstrate that the inversely derived nanostructures sticking to the inlet side of the subwavelength slits excite SPPs and guide the SPPs propagating into the metallic slits; the two streams of SPPs in the slits propagate along the two sides of the metallic slits, couple with each other, the Fabry-P\'{e}rot resonance of these two streams of SPPs is established sequentially with strengthened transmission \cite{Sorger2009,Astilean2000,Takakura2001,Yang2002,Cai2010}; at the outlet side, the resonating SPPs is scattered and radiated into free space by the inversely derived nanostructures; at last, EOT is achieved and enhanced by the inversely designed bilateral nanostructures, with resonance peaks demonstrated in the transmission spectra shown in Fig. \ref{TransSpectra}.

The effectivity of the inversely designed bilateral nanostructures on enhancing the EOT performance of periodic metallic slits can be demonstrated furthermore by the transmission spectra in Fig. \ref{TransSpectra}. There are two transmission peaks at the wavelength $526$nm and $616$nm for the periodic metallic slits without bilateral nanostructures. These transmission peaks are enhanced $3.27$ and $4.06$ times respectively by the inversely designed bilateral nanostructures with geometrical configurations shown in Fig. \ref{SlitTop}c and \ref{SlitTop}d. In Fig. \ref{TransSpectra}, the transmission peaks are presented at the specified incident wavelengths in the inverse design procedure, and the peak is red shifted along with the increase of the incident wavelength. Therefore, the derived nanostructures result in the resonant EOT performance at the specified incident wavelength, i.e. the nanostructures derived by the inverse design method are resonant for EOT of periodic metallic slits.  Sequentially, this inverse design method can be used to control the presentation of the transmission peak at a desired frequency, by specifying the incident wavelength corresponding to the desired frequency in the inverse design procedure.

\begin{figure}[htbp]
  \centering
  \subfigure[$\lambda=350\mathrm{nm}$]
  {\includegraphics[height=0.22\columnwidth]{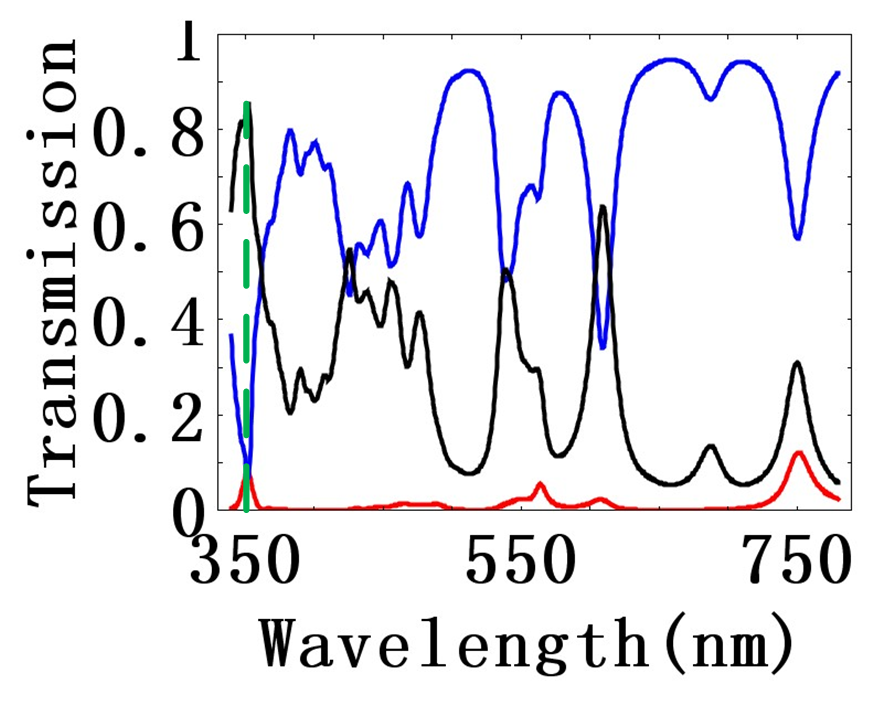}}\hspace{2em}
  \subfigure[$\lambda=440\mathrm{nm}$]
  {\includegraphics[height=0.22\columnwidth]{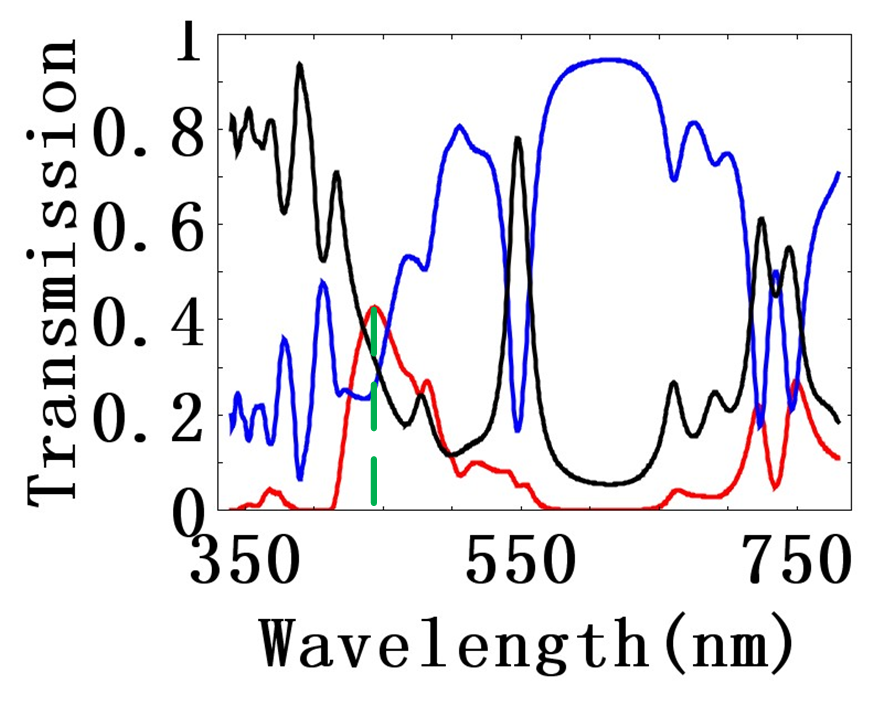}}\hspace{2em}
  \subfigure[$\lambda=526\mathrm{nm}$]
  {\includegraphics[height=0.22\columnwidth]{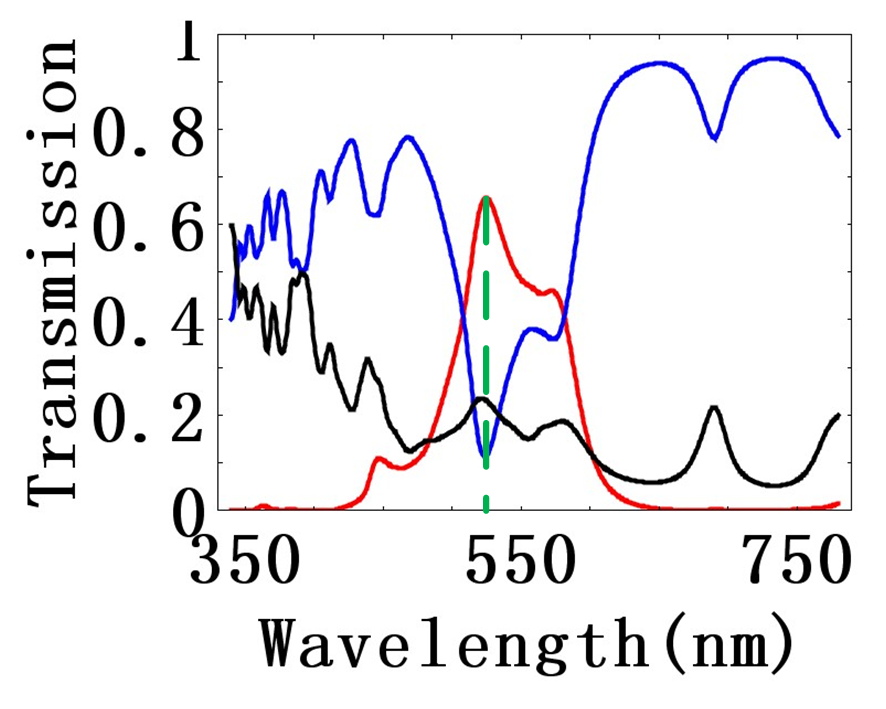}}\\
  \subfigure[$\lambda=616\mathrm{nm}$]
  {\includegraphics[height=0.22\columnwidth]{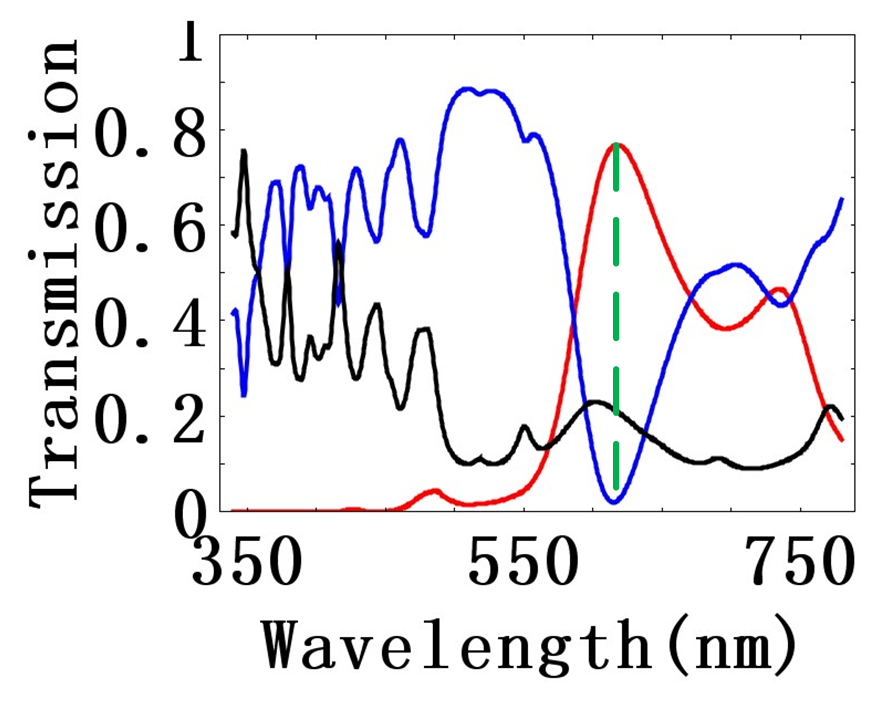}}\hspace{2em}
  \subfigure[$\lambda=710\mathrm{nm}$]
  {\includegraphics[height=0.22\columnwidth]{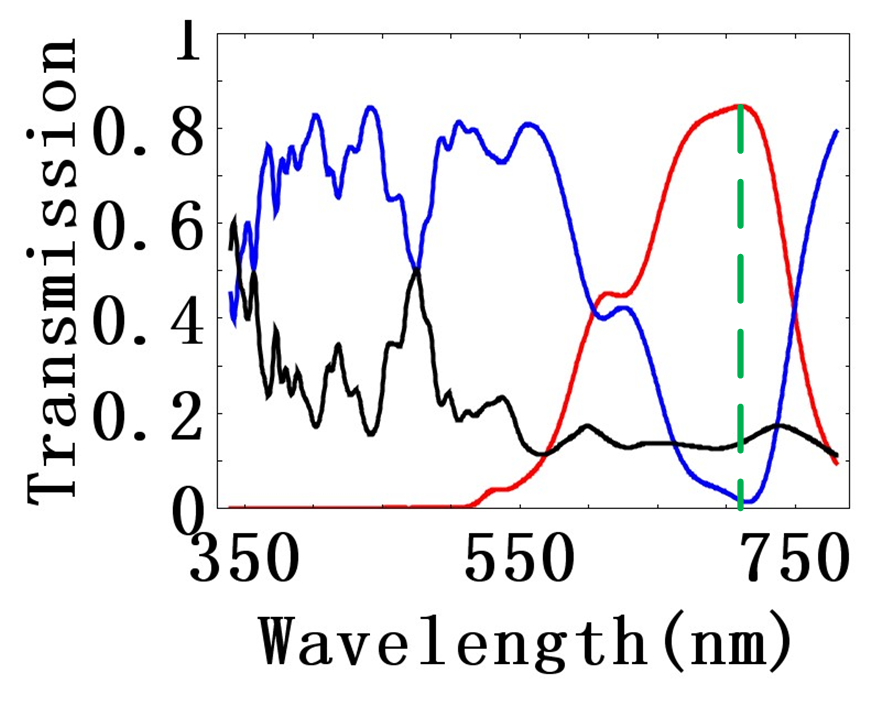}}\hspace{2em}
  \subfigure[$\lambda=770\mathrm{nm}$]
  {\includegraphics[height=0.22\columnwidth]{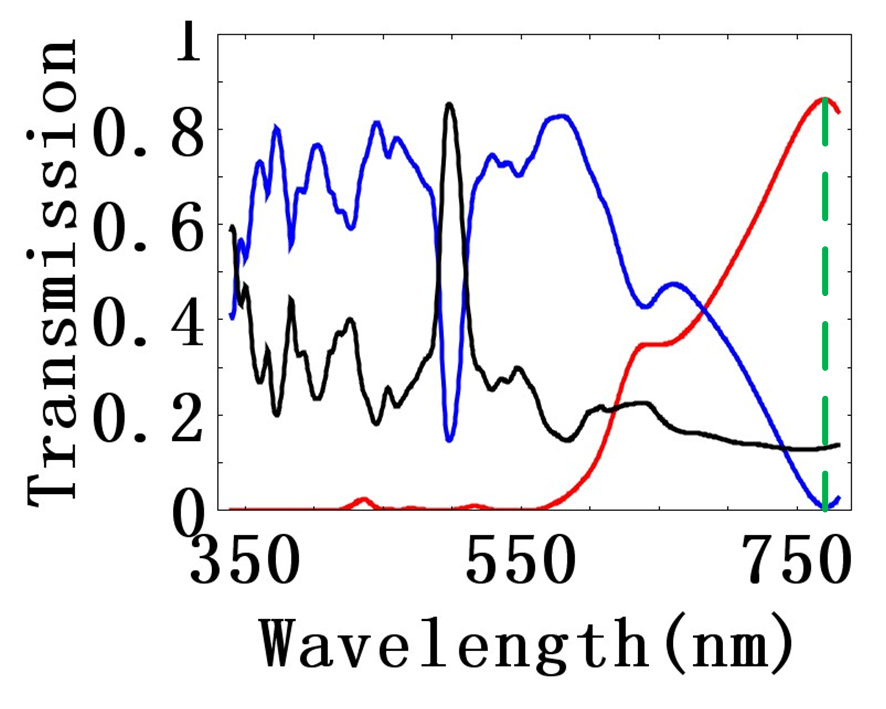}}\\
  \includegraphics[height=0.04\columnwidth]{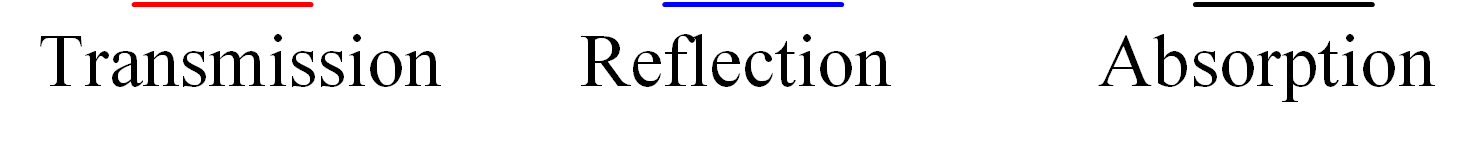}
  \caption{Transmission, reflection and absorption spectra of the periodic metallic slits respectively corresponding to the geometrical configurations shown in Fig. \ref{SlitTop}a$\sim$f, where large absorption is presented in the short-wavelength region, and relatively low absorption is presented as the red sift of the wavelength.}\label{SpectraTRA}
\end{figure}

The transmission spectra in Fig. \ref{TransSpectra} also demonstrates that the transmission peak is raised along with the red shift of the incident wavelength. This is mainly caused by the large absorption of the photonic energy and short propagation distance of SPP in the short-wavelength region, and relatively low absorption and long propagation distance of SPP in the long-wavelength region. And this can be confirmed by the comparison of transmission, reflection and absorption spectra of the periodic metallic slits with inversely designed bilateral nanostructures (Fig. \ref{SpectraTRA}). The absorption spectra in Fig. \ref{SpectraTRA} also demonstrate that the periodic metallic slits with inversely derived bilateral nanostructures have resonant light absorption performance, e.g. $95\%$ of the photonic energy is absorbed by the slits shown in Fig. \ref{SlitTop}b at the incident wavelength 390nm.

\begin{figure}[!htbp]
  \centering
  \subfigure[Metallic slits and the corresponding field distribution]
  {\includegraphics[height=0.34\columnwidth]{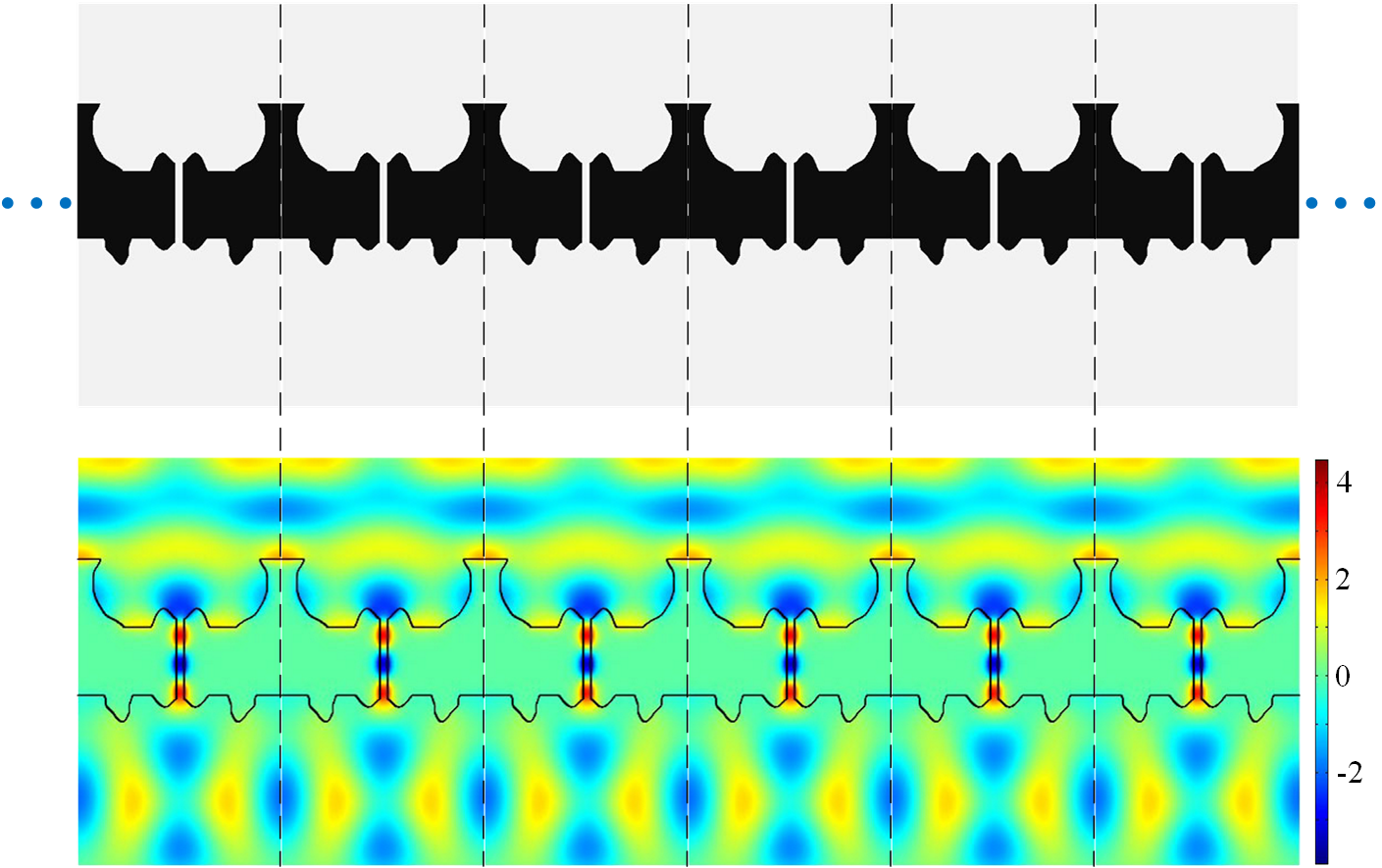}}
  \subfigure[Transmission spectra]
  {\includegraphics[height=0.34\columnwidth]{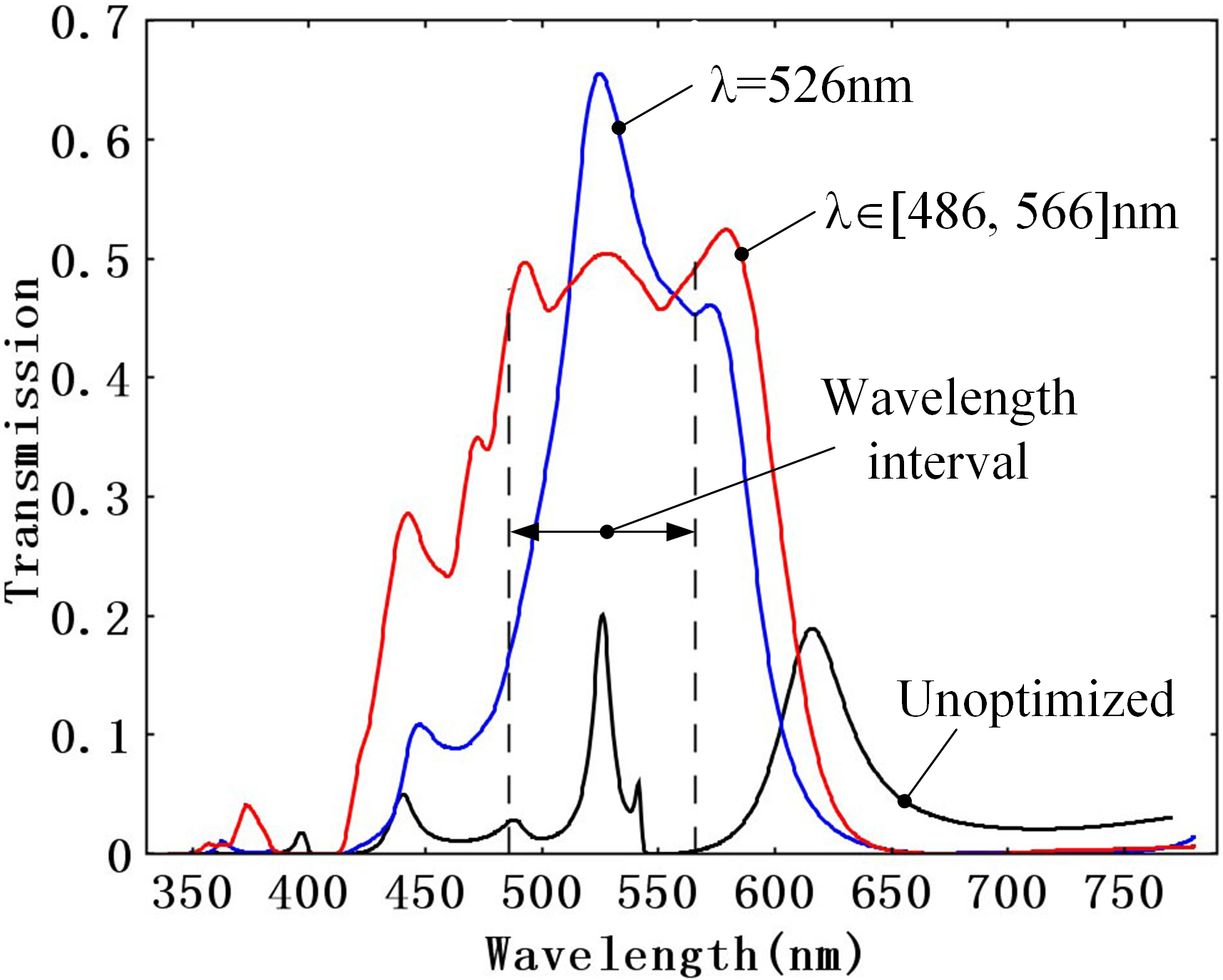}}
  \caption{(a) Periodic metallic slits with inversely designed bilateral nanostructures for the TM wave with incident wavelength in the range from $486$ to $566$nm, and the corresponding magnetic field distribution in the derived metallic slit configuration corresponding to the minimum transmissivity at the wavelength $504$nm in the prescribed wavelength range; (b) transmission spectra of the derived metallic slit configuration, where the spectra of the periodic metallic slits with bilateral nanostructures inversely designed at the central wavelength of the wavelength range and that of the periodic metallic slits without bilateral nanostructures are included.}\label{Slit486_566}
\end{figure}

\begin{figure}[!htbp]
  \centering
  \subfigure[Metallic slits and the corresponding field distribution]
  {\includegraphics[height=0.34\columnwidth]{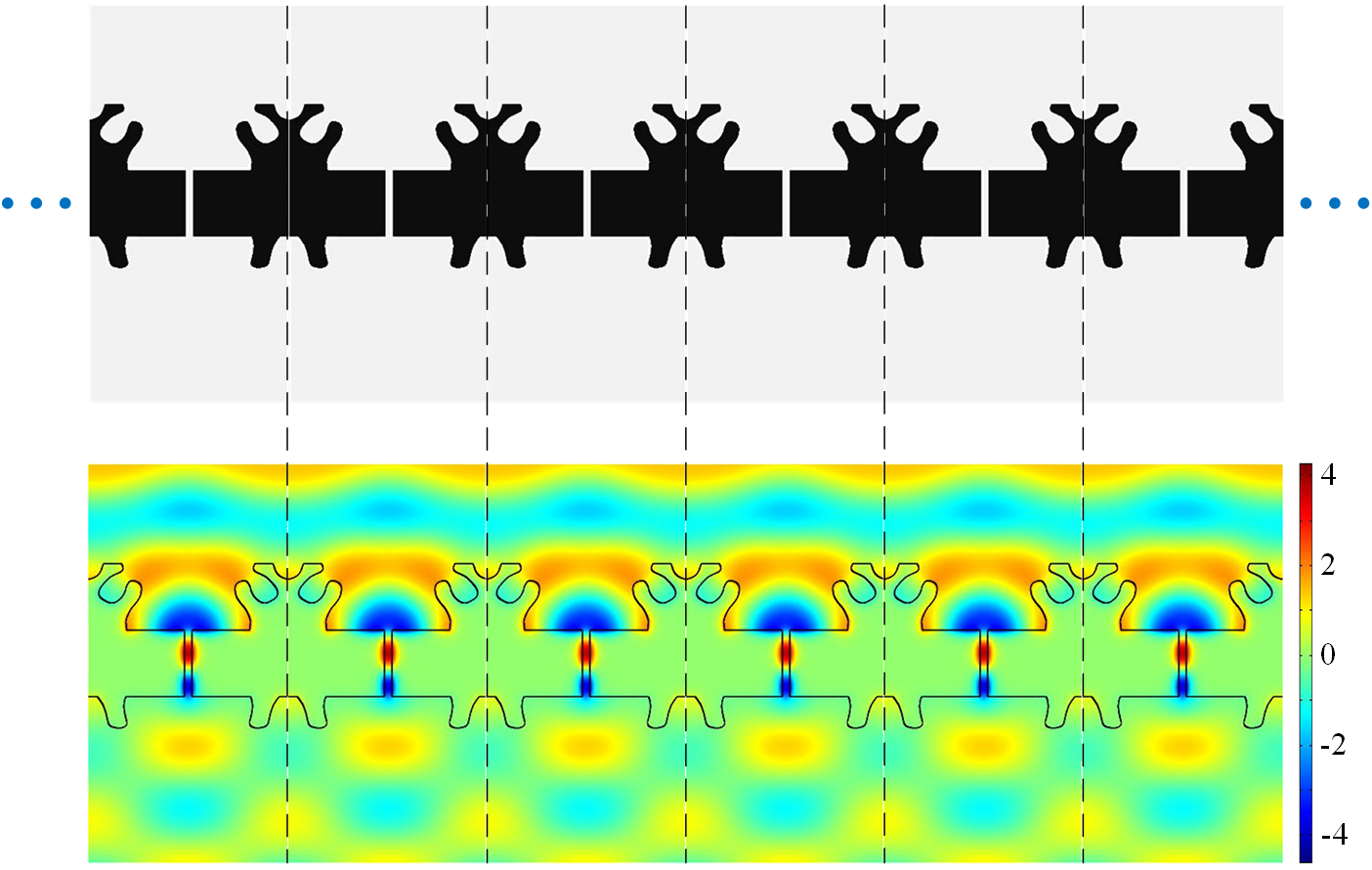}}
  \subfigure[Transmission spectra]
  {\includegraphics[height=0.34\columnwidth]{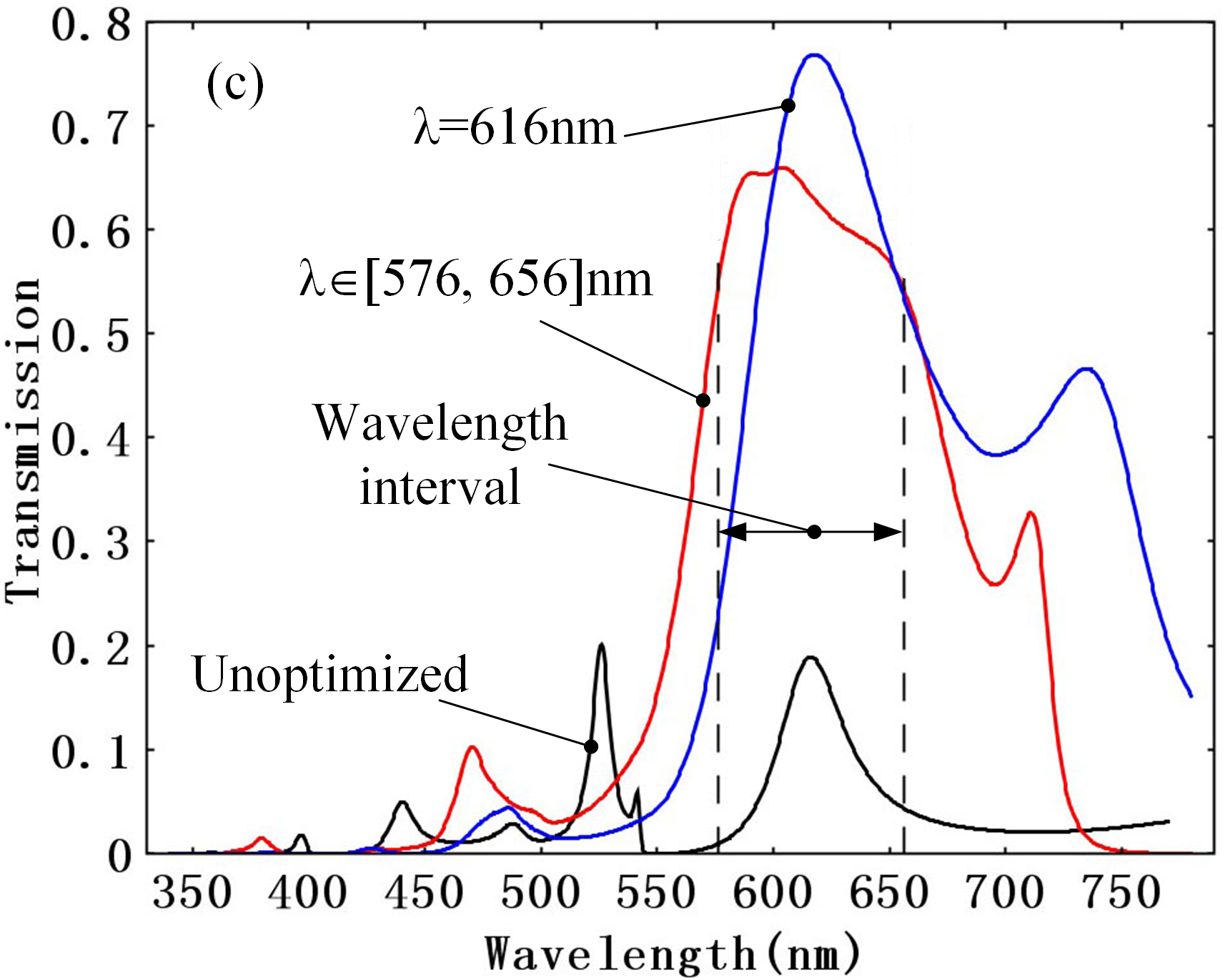}}
  \caption{(a) Periodic metallic slits with inversely designed bilateral nanostructures for the TM wave with incident wavelength in the range from $576$ to $656$nm, and the corresponding magnetic field distribution in the derived metallic slit configuration corresponding to the minimum transmissivity at the wavelength $576$nm in the prescribed wavelength range; (b) transmission spectra of the derived metallic slit configuration, where the spectra of the periodic metallic slits with bilateral nanostructures inversely designed at the central wavelength of the wavelength range and that of the periodic metallic slits without bilateral nanostructures are included.}\label{Slit576_656}
\end{figure}

The discussed inverse design method can be extended to enlarge the EOT bandwidth of the periodic metallic silts. And by enlarge the EOT bandwidth, the sensitivity of the transmissivity to the incident wavelengths can be reduced in a specified wavelength range. Then one $maximum-minimum$ inverse design objective is formulated for the bilateral nanostructures of the periodic metallic slits at a central incident wavelength with specified wavelength perturbation range
\begin{equation}\label{MaxMin}
    \max_{\rho\in\left[0,1\right]}\left\{\min_{\lambda\in\left[\lambda_0-{\theta\over2}, \lambda_0+{\theta\over2}\right]} T_r\left(\lambda\right)\right\}
\end{equation}
where $\lambda_0$ is the central incident wavelength; $\theta$ is the support size of the wavelength perturbation range. By respectively setting the central incident wavelength and support size to be $526$ and $80$nm, the geometrical configuration of the periodic metallic slits is derived as shown in Fig. \ref{Slit486_566}a, where the magnetic field distribution corresponding to the minimum transmissivity at the wavelength $504$nm is included; and the transmission spectra of the derived periodic metallic slits is shown in Fig. \ref{Slit486_566}b. When the central incident wavelength is changed to be $616$nm, the geometrical configuration shown in Fig. \ref{Slit576_656}a is derived for the TM wave with incident wavelength in the range from $576$ to $656$nm, where the magnetic field distribution corresponding to the minimum transmissivity at the wavelength $576$nm is included; and the corresponding transmission spectra is shown in Fig. \ref{Slit576_656}b. The transmission spectra in Fig. \ref{Slit486_566}b and \ref{Slit576_656}b, demonstrate that EOT is controlled to be less sensitive to the incident wavelength with the cost of decreasing transmissivity and it is feasible to enlarge the EOT bandwidth of subwavelength structures using the discussed inverse design method.

\section{Conclusion}

This paper has presented inversely determining the resonant configuration of the bilateral nanostructures for periodic metallic slits with extraordinary optical transmission performance. The topology optimization approach is utilized to implement the inverse design procedure. Several geometrical configurations of the bilateral nanostructures are derived for periodic metallic slits. The resonant performance of the derived nanostructures are demonstrated by the transmission spectra, where the transmission peak is presented at the specified wavelength in the inverse design procedure. This provides an approach to control the red or blue shift of the transmission peak or localize the resonant performance at a desired frequency, by specifying the desired incident wavelength in the inverse design procedure. The inverse design method is extended to make the periodic metallic slits to be less sensitive to the incident wavelength. This research can be further extended to inversely find the resonant subwavelength structures for extraordinary optical absorption and other surface palsmon polariton based photonic devices. The inverse design of three dimensional apertures for extraordinary optical transmission will be investigated in our future researches.

\section*{Acknowledgement}
This work is supported by the National Natural Science Foundation of China (No. 51405465), the National High Technology Program of China (No. 2015AA042604), Science and Technology Development Plan of Jilin Province and Changchun City (No. 20170312027ZG, No. 15SS12).

\end{document}